\documentclass{ifacconf}
\usepackage{graphicx}      %
\usepackage{natbib}        %
\usepackage{tikz}

\usepackage{amsmath,amssymb,amsfonts}
\usepackage{mdframed}
\usepackage{xcolor}
\usepackage{url}

\usepackage{dsfont}

\usepackage{dirtytalk}

\usepackage{endnotes}

\newcommand{\rev}[1]{\textcolor{blue}{#1}}

\definecolor{cadetblue}{rgb}{0.33, 0.41, 0.58}

\DeclareMathOperator{\argmin}{\mathrm{argmin}}

\DeclareMathOperator{\EV}{\mathds{E}}

\DeclareMathAlphabet{\doublestruck}{U}{BOONDOX-ds}{m}{n}

\newcommand{\N}[0]{\mathbb{N}}

\newcommand{\Rp}[0]{\mathbb{R}_{>0}}

\newcommand{\Acal}[0]{\mathcal{A}}

\newcommand{\Ccal}[0]{\mathcal{C}}

\newcommand{\Hcal}[0]{\mathcal{H}}

\newcommand{\Kcal}[0]{\mathcal{K}}

\newcommand{\Pcal}[0]{\mathcal{P}}

\newcommand{\Tcal}[0]{\mathcal{T}}
\newcommand{\Ucal}[0]{\mathcal{U}}

\newcommand{\Xcal}[0]{\mathcal{X}}

\newcommand{\Wcal}[0]{\mathcal{W}}

\usepackage{xstring}

\newcommand{\Rr}[0]{R_\mathrm{r}}
\newcommand{\Rno}[0]{R_\mathrm{n}}

\newcommand{\Jeff}[0]{J_\mathrm{Eff}}

\makeatletter
\providecommand{\href}[2]{#2}
\ifdefined\pdfstartlink
  \renewcommand{\href}[2]{%
    \leavevmode
    \pdfstartlink attr{/Border[0 0 0]} user{/Subtype /Link /A << /S /URI /URI (\pdfescapestring{#1}) >>}%
    #2%
    \pdfendlink
  }
\fi
\makeatother

\definecolor{lightergray}{gray}{0.9}
\mdfdefinestyle{callout}{%
	linecolor=black,
	linewidth=1pt,
	roundcorner=0pt,
	innertopmargin=4pt,
	innerbottommargin=4pt,
	innerrightmargin=5pt,
	innerleftmargin=5pt,
	leftmargin=0pt,
	rightmargin=0pt,
	backgroundcolor=lightergray
}

\providecommand{\Jeff}[0]{J_\mathrm{Eff}}

\begin{document}

\begin{frontmatter}

\title{Token Economy Design for Fair and Efficient Highway Congestion Management with Express Lanes}

\author[First]{Leonardo Pedroso} 
\author[First]{Juan Pablo Bertucci} 
\author[First]{W.P.M.H. (Maurice) Heemels}
\author[First]{Mauro Salazar}

\address[First]{Control Systems Technology section, Department of Mechanical
Engineering, Eindhoven University of Technology, The Netherlands (e-mail: \{l.pedroso,j.p.bertucci,m.heemels,m.r.u.salazar\}@tue.nl)}

\begin{abstract}  %
We study the design of a token economy for highway lane allocation that aims to improve fairness without sacrificing traffic efficiency. Motivated by the San Mateo 101 Express Lanes Project, we consider a setting in which high-occupancy vehicles have unrestricted access to an express lane, while the remaining users can alternate between regular and express lanes by earning and spending nonmonetary tokens. We model the resulting interaction as a finite-population dynamic congestion game with heterogeneous time preferences and limited-information evolutionary policy revisions. Building on a mean-field approximation, we derive token prices that enforce the system-optimal lane split while inducing fairness over time through turn-taking. The scheme is evaluated in a microscopic traffic simulation with real-world demand data. The results show that the proposed prices yield nearly the same average travel time as a baseline scenario in which no lane is reserved as an express lane, while substantially reducing urgency-weighted perceived travel time. These findings highlight token economies as a promising alternative to monetary congestion pricing for fairer management of scarce road capacity.\vspace{-0.2cm}
\end{abstract}

\begin{keyword}
Intelligent transportation systems, Integrated traffic management, Stochastic control and game theory, Mechanism design, Congestion games, Token economy
\end{keyword}

\end{frontmatter}
\renewcommand{\rev}[1]{#1}

\section{Introduction}\label{sec:introduction}

\vspace{-0.6cm}

Recent progress in connectivity and the Internet of Things has fueled the growth of sharing economies in modern society. In these settings, users compete for access to shared resources such as mobility infrastructures, cloud computing services, and electrical power. Harnessing the full potential of these resources therefore requires carefully designed allocation rules. This is a mechanism design problem~\citep{ChremosMalikopoulos2024}, and it has recently been identified as an urgent societal challenge for the control community~\citep{AnnaswamyJohanssonEtAl2023alt}.

A well-known instance is highway vehicular traffic, where individual route choices can lead to congestion and reduce overall efficiency. One common approach to mitigating congestion is to reserve specific road lanes, such as high-occupancy vehicle or carpool lanes, for vehicles carrying multiple passengers. This strategy aims to increase person throughput by incentivizing ride-sharing. However, recent studies suggest that there is room to improve the efficiency of traditional carpool lane policies~\citep{LapardhajaEtAl2021}. To leverage this opportunity, in 2023, so-called express lanes were set up along a 22-mile stretch in both directions on U.S.\ 101 from the San Mateo County/Santa Clara County line to I-380 in South San Francisco~\citep{Caltrans2026}. In contrast to traditional carpool lanes, express lanes can be used by both (i)~vehicles with three or more occupants and (ii)~vehicles with two or fewer occupants upon payment of a monetary toll.

While this approach increases throughput, it introduces fairness concerns: higher-income individuals can afford solo access to faster lanes, creating a socioeconomic bias. This fairness shortfall has been noted in the San Mateo 101 Express Lanes (SM 101) Project, which motivated the introduction of an equity program~\citep{SMCTA2023}. This program leveraged a credit-based congestion pricing scheme, in which low-income residents are given a one-time $100$~USD credit to use the express lane. In~\cite{JalotaEtAl2023}, credit-based congestion pricing schemes are judiciously designed using this project as a case study.

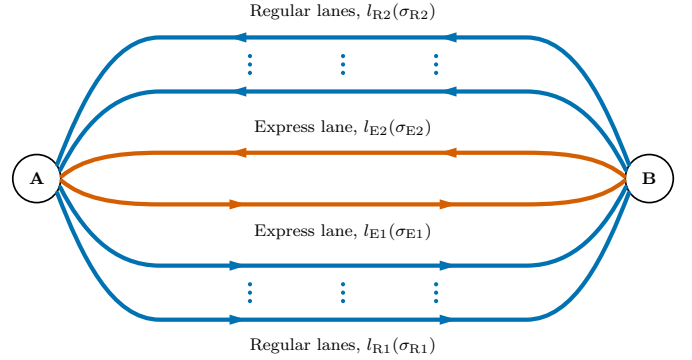
\begin{figure}[t]
	\centering
	\begingroup
	\definecolor{resourcesregular}{RGB}{0,114,178}
	\definecolor{resourcesexpress}{RGB}{213,94,0}
	\resizebox{\linewidth}{!}{%
	\begin{tikzpicture}[
		x=1cm,
		y=1cm,
		line cap=round,
		line join=round,
		resources location/.style={
			circle,
			draw=black,
			fill=white,
			line width=0.8pt,
			minimum size=8mm,
			inner sep=0pt,
			font=\small\bfseries
		},
		resources regular lane/.style={
			draw=resourcesregular,
			line width=2pt
		},
		resources express lane/.style={
			draw=resourcesexpress,
			line width=2pt
		},
		resources label/.style={
			font=\small,
			inner sep=0pt,
			align=center
		}
	]
		\node[resources location] (resources-a) at (0,0) {A};
		\node[resources location] (resources-b) at (10.2,0) {B};

		\draw[resources express lane]
			(resources-b.175)
			.. controls (9.45,0.32) and (8.95,0.43) .. (8.15,0.43)
			-- (2.05,0.43)
			.. controls (1.25,0.43) and (0.75,0.32) .. (resources-a.5);
		\draw[resources regular lane]
			(resources-b.160)
			.. controls (9.45,0.85) and (8.95,1.45) .. (8.15,1.45)
			-- (2.05,1.45)
			.. controls (1.25,1.45) and (0.75,0.85) .. (resources-a.20);
		\draw[resources regular lane]
			(resources-b.145)
			.. controls (9.45,1.30) and (8.95,2.35) .. (8.15,2.35)
			-- (2.05,2.35)
			.. controls (1.25,2.35) and (0.75,1.30) .. (resources-a.35);

		\draw[resources express lane]
			(resources-a.355)
			.. controls (0.75,-0.32) and (1.25,-0.43) .. (2.05,-0.43)
			-- (8.15,-0.43)
			.. controls (8.95,-0.43) and (9.45,-0.32) .. (resources-b.185);
		\draw[resources regular lane]
			(resources-a.340)
			.. controls (0.75,-0.85) and (1.25,-1.45) .. (2.05,-1.45)
			-- (8.15,-1.45)
			.. controls (8.95,-1.45) and (9.45,-0.85) .. (resources-b.200);
		\draw[resources regular lane]
			(resources-a.325)
			.. controls (0.75,-1.30) and (1.25,-2.35) .. (2.05,-2.35)
			-- (8.15,-2.35)
			.. controls (8.95,-2.35) and (9.45,-1.30) .. (resources-b.215);

		\foreach \x in {3.35,6.85} {
			\foreach \y in {1.45,2.35} {
				\fill[resourcesregular]
					(\x-0.14,\y) -- (\x+0.14,\y+0.085) -- (\x+0.14,\y-0.085) -- cycle;
			}
			\fill[resourcesexpress]
				(\x-0.14,0.43) -- (\x+0.14,0.515) -- (\x+0.14,0.345) -- cycle;
		}
		\foreach \x in {3.35,6.85} {
			\foreach \y in {-1.45,-2.35} {
				\fill[resourcesregular]
					(\x+0.14,\y) -- (\x-0.14,\y+0.085) -- (\x-0.14,\y-0.085) -- cycle;
			}
			\fill[resourcesexpress]
				(\x+0.14,-0.43) -- (\x-0.14,-0.345) -- (\x-0.14,-0.515) -- cycle;
		}

		\foreach \x in {3.55,5.10,6.65} {
			\foreach \y in {1.76,1.90,2.04} {
				\fill[resourcesregular] (\x,\y) circle[radius=0.035];
				\fill[resourcesregular] (\x,-\y) circle[radius=0.035];
			}
		}

		\node[resources label] at (5.10,2.78)
			{Regular lanes, $l_{\mathrm{R2}}(\sigma_{\mathrm{R2}})$};
		\node[resources label] at (5.10,0.84)
			{Express lane, $l_{\mathrm{E2}}(\sigma_{\mathrm{E2}})$};
		\node[resources label] at (5.10,-0.88)
			{Express lane, $l_{\mathrm{E1}}(\sigma_{\mathrm{E1}})$};
		\node[resources label] at (5.10,-2.78)
			{Regular lanes, $l_{\mathrm{R1}}(\sigma_{\mathrm{R1}})$};
	\end{tikzpicture}
	}
	\endgroup
	\caption{Scheme of the express lane commute problem.}\label{fig:resources}
\end{figure}

Despite improving fairness, such one-time credit-based schemes still enable higher-income individuals to repeatedly use the express lane, thereby increasing congestion, whereas other users are forced to use it sporadically. Furthermore, the fairness improvement wears off over time as disadvantaged users deplete their travel credits. The goal of this paper is to propose, design, and evaluate the performance of an incentive scheme that addresses these shortcomings.

The incentive scheme relies on a \emph{token economy} that leverages repeated allocation settings to achieve both fairness and efficiency. This is achieved by incentivizing turn-taking behavior, whereby users take turns using the most desirable resources. Each user has a wallet of tokens, which cannot be traded or bought with money. Whenever they choose a resource, they pay a corresponding amount of tokens, and they are restricted to choosing resources that they can afford. Crucially, some resources have a negative price, meaning that users earn tokens by choosing them. In the context of the SM 101 Project case study, each time a user chooses the express lane, they pay a price in tokens, and each time they choose to travel only in regular lanes, they earn tokens. This induces turn-taking behavior between express and regular lanes. Moreover, users are empowered to choose when to spend tokens to travel faster, depending on their time-varying needs. For example, they can choose to use tokens when they are in a rush and save them otherwise, which substantially lowers their perceived discomfort while traveling. Other turn-taking mechanisms based on tokens have been proposed in a similar setting in \cite{ElokdaCenedeseEtAl2022}, but they require users to submit a token bid every time they wish to use the express lane, which may lead to decision fatigue.

In this paper, we model the token economy scheme on the express lane allocation problem as a finite-population dynamic game. The main contributions of this work are twofold:
\begin{itemize}
    \item We use a mean-field approximation of the finite-population token economy model together with very limited assumptions on the decision-making behavior of users to design token prices that achieve optimal efficiency and fairness.
    \item We validate the fairness and efficiency of the token economy framework in a microscopic traffic simulator with realistic demand profiles of the SM 101 Project case study. We show that this framework has the potential to reduce perceived user discomfort and average travel time through a fair mechanism.
\end{itemize}

\emph{Notation}: The set of consecutive positive integer numbers $\{1,2,\ldots, N\}$ is denoted by $[N]$. The indicator function of $a\in \Xcal$ is denoted by $\delta_a:\Xcal \to \{0,1\}$ and $\delta_a(x) = 0$ if $x\neq a$ and $\delta_a(x) = 1$ if $x = a$. The set of all Borel probability measures on a finite set $\Acal$ is denoted by $\Pcal(\Acal)$.

\section{Token Economy Model}\label{sec:model}

\vspace{-0.2cm}

In this section, we formulate a model of a token economy for the express lane commute problem. Consider two locations A and B that are connected in both directions by a stretch of highway, as depicted in Fig.~\ref{fig:resources}. First, we characterize the token economy as a dynamic game of a finite population of users. Second, we introduce an information structure and an evolutionary model to characterize how users make decisions and how they revise them as the aggregate decisions of the population evolve.

\subsection{Finite-Population Dynamic Congestion Game Model}\label{sec:finite_model}

The token economy of the express lane commute problem is modeled as follows:
\begin{itemize}
	\item \emph{Population}: Consider a finite population of $N$ users. Users are divided into classes according to their time preference to travel in each direction. Specifically, a $24$-hour timeline is divided into $C$ bins, where the duration of a bin $c\in[C]$ is denoted by $T_c$. Each user is characterized by a time-invariant preferred bin to travel in each direction. Formally, each user $i~\in [N]$ is characterized by $c^i_{\mathrm{AB}}\in [C]$ and $c^i_{\mathrm{BA}}\in [C]$. The set of players that are in bin $c\in [C]$ in direction $\mathrm{A}\rightarrow \mathrm{B}$ is denoted by $\Ccal^{AB}_c :=\{i\in [N]: c^i_{\mathrm{AB}}= c\}$. The set $\Ccal^{\mathrm{BA}}_c$ is defined analogously. Furthermore, we consider that a fraction $\eta \in (0,1)$ of all users always carries two or more passengers on board their vehicle and, thus, always has access to the express lanes. The set of such users is denoted by $\Hcal \subset [N]$.

	\item \emph{Time}: Each user uses the lanes in continuous time. Each day, users are randomly chosen to travel in each bin according to a uniform distribution to meet a known demand profile. Crucially, if a user is chosen to make an outbound trip on one day, they will also be chosen to make an inbound trip that same day. It is important to note that users do not necessarily travel daily. The start time of each user's trip is chosen uniformly at random within the bin's time interval. We denote the time of the $k$-th travel (in either direction) of a player $i\in [N]$ by a random variable (r.v.) $t^i_k$.
	
	\item \emph{Resources}: When traveling in either direction, users can choose between using two resources: \emph{regular lanes} or \emph{express lane} as depicted in Fig.~\ref{fig:resources}.

	\item \emph{States}: At each time $t$, each player $i\in [N]$ is characterized by: (i)~the amount of tokens that they possess, characterized by a r.v.\ $k^i(t)$, which is an integer in $\Kcal:=\{0,1,\ldots,\bar{k}\}$ where $\bar{k} \in \N$ is a maximum number of tokens allowed; and (ii)~a sensitivity to reward, which is characterized by a r.v.\ $w^i(t)$, which is in the finite set $\Wcal \subseteq \Rp$. The pair $(k^i(t),w^i(t))$ is called the state of user $i$ at time $t$.
	
	\item \emph{Pricing mechanism}:  We levy integer token prices for each choice of route, which are denoted by $\tau_{\mathrm{R1}}(t)$, $\tau_{\mathrm{E1}}(t)$, $\tau_{\mathrm{R2}}(t)$, and $\tau_{\mathrm{E2}}(t)$ that may depend on time $t$. We assume that $\tau_{\mathrm{R1}}(t) \leq 0$  and $\tau_{\mathrm{R2}}(t) \leq 0$ so that every user can always afford at least one action.

	\item \emph{Actions}: When a user $i \in [N]$ enters the highway at time $t$ at $\mathrm{A}$ (at $\mathrm{B}$), they must choose an action $a^i(t)\in \Acal_1 := \{\mathrm{R1},\mathrm{E1}\}$ ($a^i(t)\in \Acal_2 := \{\mathrm{R2},\mathrm{E2}\}$) that they can afford, i.e., an action in $\Acal_1(k^i(t),t):=  \{a\in \Acal_1 : \tau_a(t) \leq k^i(t)\}$ ($\Acal_2(k^i(t),t)$, which is defined analogously).

	\item \emph{State transitions}: When a player $i\in [N]$ commutes and chooses a route $a \in \{\mathrm{R1},\mathrm{E1},\mathrm{R2},\mathrm{E2}\}$, their state evolves according to a Markov transition kernel. Specifically, their amount of tokens evolves according to a deterministic Markov kernel $\phi_\Kcal^a: \Kcal \to  \Pcal(\Kcal)$ defined as
	\begin{equation*}\label{eq:kernel_K}
		\phi^a_\Kcal(k) = 	\begin{cases}
		\delta_{k-\tau_a(t)}(\cdot), & k-\tau_a(t) < \bar{k}\\
		\delta_{\bar{k}}(\cdot), & k-\tau_a(t) \geq \bar{k},
		\end{cases}
	\end{equation*}
	and the sensitivity to reward evolves according to a time-invariant Markov kernel denoted by $\phi_\Wcal: \Wcal \to  \Pcal(\Wcal)$.

	\item \emph{Single-stage reward}: All resources are associated with latency functions, which are denoted by $l_{\mathrm{R1}}(\sigma_{\mathrm{R1}}(t))$, $l_{\mathrm{E1}}(\sigma_{\mathrm{E1}}(t))$, $l_{\mathrm{R2}}(\sigma_{\mathrm{R2}}(t))$, and $l_{\mathrm{E2}}(\sigma_{\mathrm{E2}}(t))$, that characterize the travel time on that route depending on the flow of vehicles on that resource. We assume that the latency functions are strictly increasing.
	
	\item \emph{Payoff}: The payoff of a user $i\in [N]$ is modeled as the long-time average reward weighted by individual reward sensitivities, which is given by
	\begin{equation}\label{eq:payoff}
		\!\!\!J^i\!:= \!-\lim_{T\to \infty}\frac{1}{T}\EV\left[\sum_{k=1}^T w^i(t^i_k)l_{a^i(t^i_k)}({\sigma_{a^i(t^i_k)}}(t^i_k))\right]\!.
	\end{equation}
 \end{itemize}

\subsection{Evolutionary Decision Model}\label{sec:ev_dynamics}

\vspace{-0.2cm}

The information structure of the token economy model characterizes the information that each user has access to during the game. Given the information available to a user, they choose one route. We call the map from the information known by a user to their choice of action a \emph{policy}. We model the information structure of the token economy game as:
\begin{itemize}
	\item Each user \emph{does not have access to aggregate information} of the distributions of the amount of tokens or sensitivity to reward of other users. Indeed, their amount of tokens is tracked by the operator of the resources but not disclosed publicly, and their sensitivity to reward is private.
	\item Each user does not keep a history of their past token amounts, sensitivities to reward, or past actions. At each travel instant, a user knows only their current amount of tokens and sensitivity to reward.
	\item Each user cannot predict how payoffs will evolve in the future. When they choose a map from information to a decision, they plan to use it forever.
\end{itemize}
Policies that are consistent with the above information structure are said to be \emph{oblivious}, \emph{Markov}, and \emph{stationary}, respectively. Such policies are formally characterized, for each direction, by maps $u_1:\Kcal \times \Wcal \to \{\mathrm{R1},\mathrm{E1}\}$ and $u_2:\Kcal \times \Wcal \to \{\mathrm{R2},\mathrm{E2}\}$ from the state of the user to a deterministic route choice. Furthermore, since the express lanes will always be more convenient to use than the normal lanes, we assume that if a user chooses the express lane with a token amount $k \in \Kcal$ and sensitivity $w\in \Wcal$, then they will also choose the express lane with any token amount $k^\prime>k$ and sensitivity $w\in \Wcal$. As a result, the set of policies for each direction and each time bin $c \in [C]$ is characterized by 
\begin{equation*}
	\begin{split}
	&\Ucal_1^c \!:= \!\left\{ u\!:\!\Kcal\!\times\! \Wcal \!\to \{\mathrm{R1},\mathrm{E1}\} \;\!|\;\! (u(k,w) = \mathrm{R1} \;\forall k < \tau^{c,\max}_{E1}) \land \right. \\
	&\!(\forall k,k^\prime \!\!\in\! \Kcal \!:\! 
	 \left. k^\prime \!\!>\! k \!\land\! u(k,w) \!= \!\mathrm{E1} \!\!\implies\!\!  u(k^\prime\!,w) \!= \!\mathrm{E1} ) ,\forall w \! \in \! \Wcal  \right\}\!,
	\end{split}
\end{equation*}
where $\tau^{c,\max}_{E1}$ is the maximum price of the express lane in the time bin $c\in [C]$ and $\Ucal_2^c$ is defined analogously.

Thus far we have modeled the way a user chooses an action depending on their amount of tokens by making use of the concept of a policy. However, we have not considered how a user chooses which policy to use and how they may switch policies as aggregate decisions change with time. We address this modeling consideration by making use of an evolutionary model.

Evolutionary models have been thoroughly analyzed for static games, where the rewards do not depend on the evolution of an individual state of each user (e.g., \citet{Sandholm2010}). Those settings are also known as population games. The token economy model presented in this section is dynamic; therefore, the proposed evolutionary model is based on a dynamic-game version of an evolutionary model proposed in~\cite{PedrosoAgazziEtAl2025MFGAvg,PedrosoAgazziEtAl2025MFGAvgII}.

According to the evolutionary model, each user $i\!\in\! [N]$ uses a deterministic policy $u_1^i(t) \!\in\! \Ucal_1^{c^i_{\mathrm{AB}}}\!$ and $u_2^i(t) \!\in\! \Ucal_2^{c^i_\mathrm{BA}}\!$ for each direction at time $t$. During the game, users, who have access to very limited information, will revise their policy choice and possibly switch to another policy in a way that reflects the \emph{inertia} and \emph{myopia} observed in real-life behavior. For the sake of conciseness, we do not formally describe the evolutionary revision dynamics herein. For more detail, the interested reader is referred to \citet{PedrosoAgazziEtAl2025MFGAvg}. 

Coarsely, the evolutionary model is characterized by
\begin{itemize}
	\item \emph{Time}: Each user makes policy revisions in continuous time. Each user is equipped with a Poisson clock with rate $R_r$, which is significantly lower than the rate of traveling. Each time a user's revision clock rings, they revise the policy that they are using. The clocks of different players are independent and have the same rate. Revision clocks of users are independent of traveling events.
	
	\item \emph{Policy transitions}: Upon a revision opportunity of a player, their policy choice evolves according to a \emph{revision protocol}. A revision protocol is a map from the current policy of the user to a probability distribution of policies that the user may switch to. The revision protocols are characterized analogously to the literature of evolutionary dynamics for static games \citep{Sandholm2010}. Therein, physically meaningful families of revision protocols are defined, such as imitative and pairwise comparison protocols, which are the families used in this paper due to their very limited information requirements. 
\end{itemize}

These revision protocols rely on very simple but meaningful rules to describe the behavior of a user. On the one hand, following an imitative revision protocol, when a user is given a revision opportunity, they choose another user at random and imitate their policy with a probability that depends on its payoff. On the other hand, following a pairwise comparison protocol, when a user is given a revision opportunity, they choose an available policy at random and switch with nonzero probability if it strictly increases the user's current payoff. Notice that the information requirements of these families are very mild, since they require knowing, at most, the payoff of the current policy and a randomly chosen policy only.

\subsection{Performance Metrics}\label{sec:metrics}

\vspace{-0.2cm}

In this section, we formalize the concepts of efficiency and fairness in the context of this problem.

First, from the macroscopic system-level perspective, we consider that the efficiency metric associated with the aggregate use of the resources that we desire to minimize is the average travel time. This metric is a function of the normalized rate at which users use the resources defined by 
\begin{equation}\label{eq:efficiency}
	\Jeff(t) = \sum_{a\in \Acal_1 \cup \Acal_2} \sigma_a(t)l_a(\sigma_a(t)).
\end{equation}

Second, the most fundamental fairness consideration in a token economy is that \emph{users with the same needs and the same impact on congestion have the same average reward \rev{over} time, no matter other external factors such as wealth or social status}. This fairness condition is called intra-class fairness. In the context of the token economy of the express lane commute problem, the intra-class fairness condition states that, in each direction, users with the same time preference who carry fewer than two passengers must perceive the same long-time average latency, i.e., for direction $\mathrm{A} \rightarrow \mathrm{B}$ for each time bin $c\in \rev{[C]}$ there exists $L^c_{\mathrm{AB}}$ such that for all $i\in [N]\setminus \Hcal$ with $c^i_{\mathrm{AB}} = c$ it holds that 
\rev{\begin{equation}\label{eq:intra_class_fairness}
	\!\!\lim_{T\to \infty}\EV \!\left[\frac{1}{|\Tcal_{\mathrm{AB}}^{i}(T)|}\sum_{k\in \Tcal_{\mathrm{AB}}^{i}(T)} \!\!\!l_{a^i(t^i_k)}({\sigma_{a^i(t^i_k)}}(t^i_k))\right] = L^c_{\mathrm{AB}},
\end{equation}
where $\Tcal_{\mathrm{AB}}^{i}(T) := \{ k\in [T]: a^i(t^i_k) \in \{\mathrm{R1},\mathrm{E1}\}\}$.}
The condition is analogous for direction $\mathrm{B} \rightarrow \mathrm{A}$.
Notably, the intra-class fairness condition is in line with recent equity-over-time distributive principles of justice for mobility systems \citep{QiuReyEtAl2025,SalazarBetancurEtAl2025}. 

\section{Token Economy Model Design}\label{sec:descriptive}

\vspace{-0.2cm}

In this section, we use a mean-field approximation of the token economy model presented in Section~\ref{sec:model} to design the token prices, resorting to the descriptive and prescriptive analyses in prior work \citep{PedrosoAgazziEtAl2026TokenEconomy}.

\subsection{Regularization Assumptions}

\vspace{-0.2cm}

First, a key property that significantly simplifies the analysis of the game is that the long-time average reward of a user does not depend on their initial amount of tokens. This can be achieved under an additional assumption. We consider that the operator of the resources maintains a noise Poisson clock with rate $\Rno \ll \Rr$ associated with each user. Each time a user's noise clock rings, the user is given one token unit. In this particular case, the aforementioned regularization noise model has a physical interpretation and can actually be implemented. For a detailed formal analysis of how this assumption regularizes the token economy game, see~\cite[Lemma~1]{PedrosoAgazziEtAl2026TokenEconomy}.

Second, we use a mean-field approximation of the finite-population model, where the population is infinite, each user carries infinitesimal weight, and the payoff of each user depends on the mean field rather than the individual states and actions of all the players. Crucially, the mean-field approximation has approximation guarantees w.r.t.\ the finite-population model and allows for a tractable analysis. For a detailed formal analysis, see~\cite[Theorem~3]{PedrosoAgazziEtAl2025MFGAvg}.

\subsection{Design}

\vspace{-0.2cm}

First, for any token prices, the intra-class fairness condition~\eqref{eq:intra_class_fairness} is satisfied by design, due to the turn-taking behavior it induces. For a formal treatment, see~\cite[Theorem~3]{PedrosoAgazziEtAl2026TokenEconomy}.

Second, to optimize efficiency the goal is to enforce optimal flows in each direction that minimize~\eqref{eq:efficiency} in each time bin $c\in [C]$. Specifically, for direction $A\rightarrow B$ and time bin $c\in [C]$ the goal is to enforce flows $f^{c\star}_{\mathrm{R}1}$ and $f^{c\star}_{\mathrm{E}1}$ that are characterized by
\begin{equation}\label{eq:sys_opt}
	\begin{aligned}
		f^{c\star}_{\mathrm{R}1},f^{c\star}_{\mathrm{E}1} \in \underset{f^{c}_{\mathrm{R}1},f^{c}_{\mathrm{E}1}}{\argmin}\;\; & f^{c}_{\mathrm{R}1}l_{\mathrm{R}1}(f^{c}_{\mathrm{R}1})+f^{c}_{\mathrm{E}1}l_{\mathrm{E}1}(f^{c}_{\mathrm{E}1}) \\
		\mathrm{s.t.}\quad  \quad \;
		& f^{c}_{\mathrm{R}1} + f^{c}_{\mathrm{E}1} = d^c_{\mathrm{AB}} \\
        & \rev{f^{c}_{\mathrm{E}1}} \geq \eta d^c_{\mathrm{AB}}\\
		& \rev{f^{c}_{\mathrm{R}1}} \geq 0,
	\end{aligned}
\end{equation}
where \rev{$d^c_{\mathrm{AB}}$} is the total average flow for bin $c\in [C]$. Optimal flows for direction $B \rightarrow A$, denoted by $f^{c\star}_{\mathrm{R}2}$ and $f^{c\star}_{\mathrm{E}2}$, are defined analogously. Notice that the mean-field approximation allows for noninteger flows and lets us write the high-occupancy vehicle flows as $\eta d^c_{\mathrm{AB}}$. To induce such flows we use the price design procedure proposed in~\cite[Theorem~4]{PedrosoAgazziEtAl2026TokenEconomy}, which yields
\begin{equation}\label{eq:prices}
        \begin{split}
        \tau^c_{R1} &= - \mathrm{round}\left( \alpha_0(f^{c\star}_{\mathrm{E}1}-\eta d^c_{\mathrm{AB}})/{\rev{((1-\eta) d^c_{\mathrm{AB}}})}\right)\\
        \tau^c_{E1}  &=   \mathrm{round}\left( \alpha_0 {f^{c\star}_{\mathrm{R}1}}/{\rev{((1-\eta) d^c_{\mathrm{AB}}})}\right),
        \end{split}
\end{equation}
where the noise term was dropped for simplicity and we used $\alpha = \alpha_0/(l_{\mathrm{R}1}(f^{c\star}_{\mathrm{R}1})-l_{\mathrm{E}1}(f^{c\star}_{\mathrm{E}1}))$, where $\alpha_0$ is any chosen positive real number. The optimal flows and token prices are defined analogously for direction $\mathrm{B} \rightarrow \mathrm{A}$.

\begin{remark}
    The prices~\eqref{eq:prices} will only be active during congested hours, meaning that outside congested hours every user can access the express lanes free of a token charge. Interestingly, with a fixed $\alpha_0>0$, even for realistic time-varying demand throughout the day, the prices in~\eqref{eq:prices} remain approximately constant during congested hours, which will be illustrated for the SM 101 case study in Section~\ref{sec:results}. %
\end{remark}

\vspace{-0.2cm}

\section{Case-study Results}\label{sec:results}
\vspace{-0.2cm}

The token economy scheme proposed in this paper is validated using the SM 101 Project case study. We use a realistic microscopic traffic simulation and realistic demand profiles for this purpose. Crucially, no additional infrastructure is necessary to implement the token economy beyond what is already deployed for the SM 101 Project.

\subsection{Simulation Setup}\label{sec:simulation_setup}

\vspace{-0.2cm}

We perform a microscopic traffic simulation of the US-101 corridor in San Mateo County, California, spanning from the I-380 to the SR-84 interchange. The simulation is implemented in the SUMO (Simulation of Urban Mobility) simulation package~\citep{sumo}.

\emph{Traffic Network}: The road network geometry was obtained from OpenStreetMap~\citep{OpenStreetMap} and converted into a SUMO-compatible format.
The resulting network comprises a northbound and a southbound corridor, each consisting of 52 consecutive edges. The raw network was post-processed to introduce a dedicated express lane: for each edge, the leftmost lane is designated as an express lane, which can only be used by token-paying vehicles or vehicles with \rev{three or more occupants}, while the remaining lanes are open to all vehicles.
Each direction has a maximum of 6 total lanes and a minimum of 4 total lanes. A uniform speed limit of 65 mph ($\approx 105$~km/h) is set across all edges, following local speed limits.

\emph{Traffic Demand}: 
Demand profiles are derived from two real-world data sources: the Caltrans Annual Average Daily Traffic (AADT) report for 2023, and hourly sensor data from the Caltrans Performance Measurement System (PeMS) \citep{PEMS} for the US-101 corridor in District~4, San Mateo County. The PeMS data provide directional (northbound/southbound) hourly volumes aggregated over a representative week in May. The AADT data \rev{are} used to scale the PeMS hourly profiles, ensuring that the total simulated demand matches average real-world volumes. Following vehicle-occupancy patterns reported in U.S. travel surveys and the observed use of high-occupancy vehicle (HOV) facilities on the US~101 corridor \citep{caltrans_us101_cmcp_2020}, we assume that approximately 5\% of vehicle trips on urban segments of US~101 have three or more occupants, i.e., $\eta = 0.05$.

\emph{Users}:
Each user is assigned a commute direction (southbound or northbound) based on a morning demand split. Time is split into one-hour bins. Each user has a preferred outbound departure hour (in $[0, 15]$) and a return hour (in $[10, 24]$), which defines their class. To model driver heterogeneity, each user is \rev{assigned a speed factor, a time headway, and a driving imperfection parameter independently sampled from uniform distributions $\mathcal{U}(0.95, 1.05)$, $\mathcal{U}(1.1, 1.3)$\,s and $\mathcal{U}(0.0, 0.1)$}, respectively. Upon departure, each user decides whether to use the express lane or the regular lanes based on their current policy, as described in Section~\ref{sec:ev_dynamics}. This decision determines whether the vehicle is allowed to use the express lanes, which is enforced throughout the trip.

\emph{Simulation Loop}:
The simulation runs over $12$ weeks with a time step of $\Delta t = 1.0$\,s. An event-driven architecture processes two event types: (i)~trip events, which trigger vehicle injections and follow the traffic demand data; and (ii)~policy revision events, scheduled according to user-specific Poisson processes, during which users revise their policies. When a vehicle arrives at its destination, its travel time is recorded and used for the threshold evaluation for the next lane choice decision. To meet the demand data and add some more realism to the simulation, some trip requests occur outside the user's preferred departure and return hours.

For the sake of conciseness, further details about the simulation are omitted. For reproducibility, the code used to carry out the simulations in this paper is available in an open-access repository at \href{https://github.com/fish-tue/express-lane-token-economy}{github.com/fish-tue/express-lane-token-economy}.

\subsection{Design}

The first step is to fit the latency functions for each of the lanes in each direction, using simulated travel-time data obtained from simulations with random allocation to express lanes proportional to their capacity. The fit is performed for all lanes and then split between the express and regular lanes proportionally to their bottleneck capacities. Data points and fitted latency functions are omitted for conciseness.

Then, the optimization problem~\eqref{eq:sys_opt} is solved for each time bin and each direction. The corresponding token prices are computed according to~\eqref{eq:prices}. During congested hours (defined as periods with at least $70\%$ of peak demand, which correspond to 5am-9pm) the token economy prices are approximately constant and equal to $\tau_{\mathrm{R}1} = \tau_{\mathrm{R}2} = -2$ and $\tau_{\mathrm{E}1} = \tau_{\mathrm{E}2} = 7$. This is to be expected since the express lane has approximately the same capacity as a single regular lane.

\subsection{Results}

\begin{figure}
    \centering
    \includegraphics[width=.9\linewidth,trim={0.4cm 0.4cm 0.3cm 0.4cm},clip]{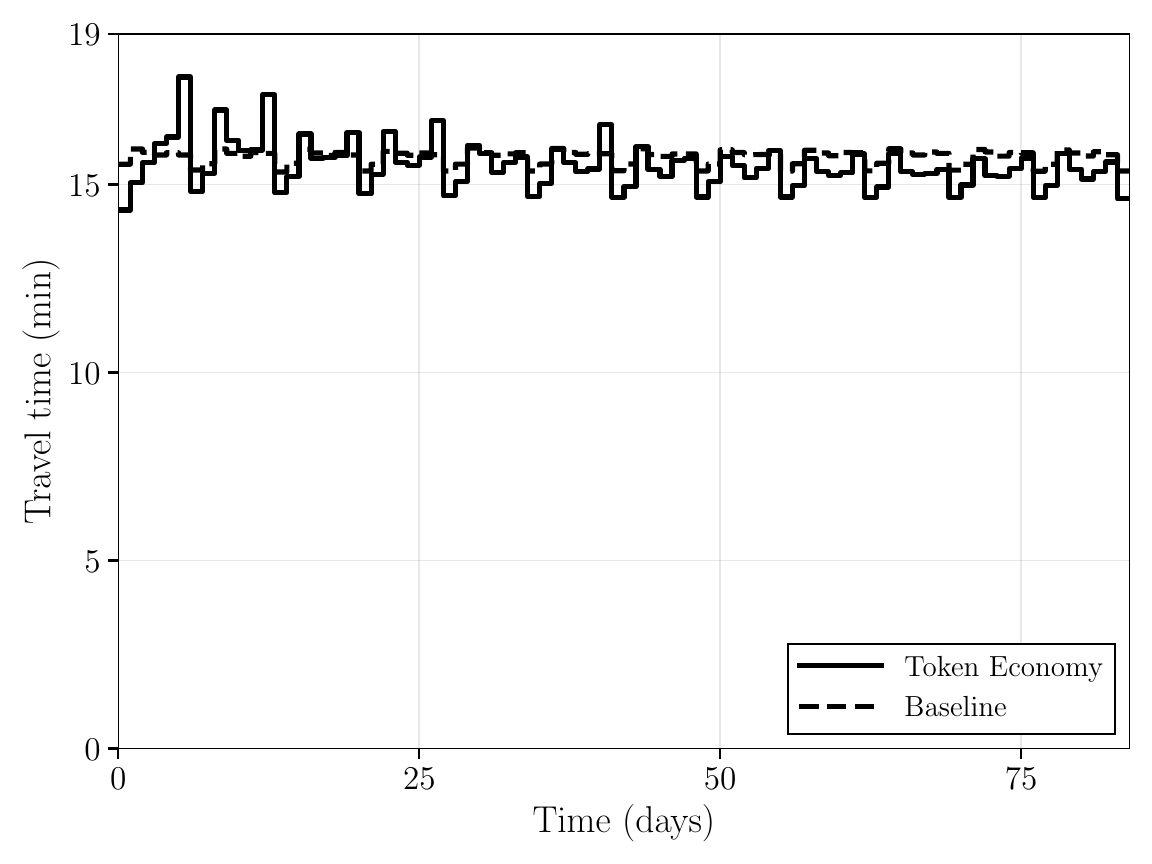}
    \caption{Evolution of overall average daily travel time.}\label{fig:efficiency}
\end{figure}

\begin{figure}
    \centering
    \includegraphics[width=.9\linewidth,trim={0.4cm 0.4cm 0.3cm 0.4cm},clip]{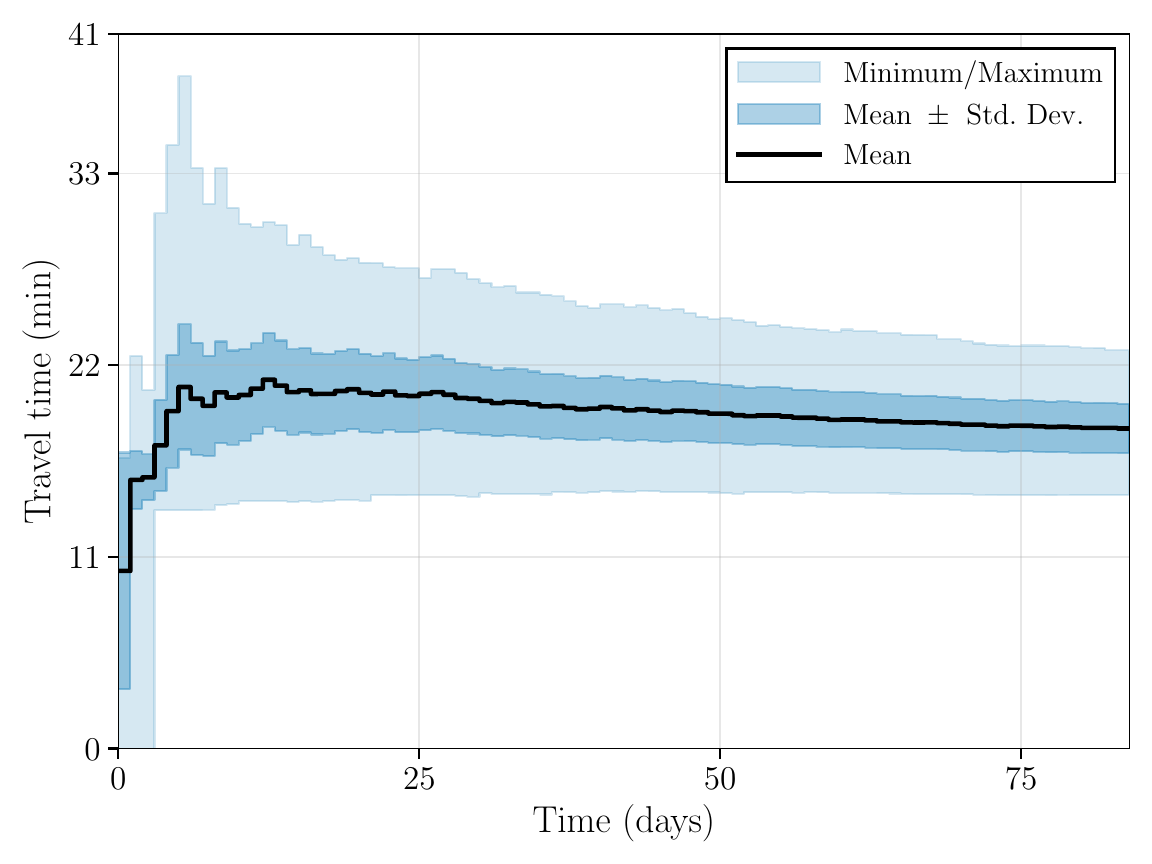}
    \caption{Distribution of average travel time on outbound trips for users on the northbound direction with an inbound trip preference between 6pm and 7pm.}\label{fig:fairness}
\end{figure}

\begin{figure}
    \centering
    \includegraphics[width=.9\linewidth,trim={0.4cm 0.4cm 0.3cm 0.4cm},clip]{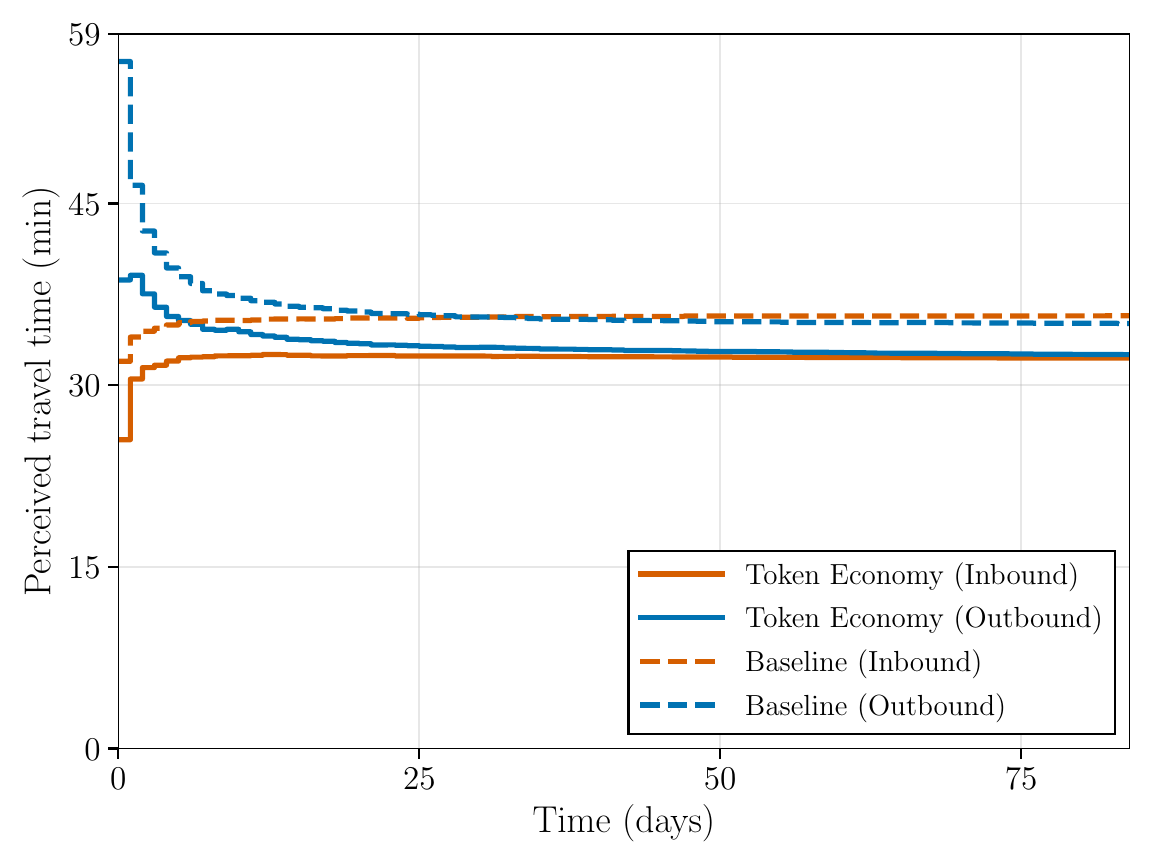}
    \caption{Evolution of urgency-weighted perceived travel time \rev{for users whose outbound commute is northbound.}}\label{fig:perceived}
\end{figure}

Fig.~\ref{fig:efficiency} depicts the evolution of the daily average travel time across all lanes and directions. Fig.~\ref{fig:fairness} depicts the average, standard deviation bounds, and extrema of travel time for all users on the northbound direction with an inbound trip preference between 6pm and 7pm. Fig.~\ref{fig:perceived} depicts the evolution of the average travel time weighted by the users' urgency, i.e., a finite-population sample of the argument of the expectation in~\eqref{eq:payoff}. The results in Figs.~\ref{fig:efficiency} and~\ref{fig:perceived} are also compared against a baseline scenario, shown with dashed lines, in which there is no express lane, i.e., all available lanes are regular lanes and no token economy is in place.

First, note that the overall average travel time is nearly identical between the scenarios with and without an express lane. Indeed, the average travel time of the baseline scenario is $15.7$ min, whereas it is $15.5$ min for the token economy scenario. This is expected since the express lane has approximately the same capacity as a single regular lane.

Second, the average travel time of users, for the class of users depicted in Fig.~\ref{fig:fairness}, converges to a tight interval, which arises from the turn-taking behavior induced by the token economy. One concludes that the intra-class fairness condition presented in Section~\ref{sec:metrics} is close to being satisfied. As expected, it is not exactly satisfied because the simulation is of a finite population in which users have heterogeneous driving behavior, which is not captured in the model of Section~\ref{sec:model}.

Third, we see in Fig.~\ref{fig:perceived} that because users are allowed to take turns using the express lane depending on their urgency to travel fast, the \rev{urgency-weighted} perceived travel time is lower than in the baseline scenario by roughly $9\%$.

\vspace{-0.1cm}
\section{Conclusion}\label{sec:conclusion}
\vspace{-0.2cm}

This paper proposed a token-economy mechanism for allocating access to express lanes in a way that combines congestion management with fairness over time. The main idea is to replace repeated monetary payments with nonmonetary tokens that users earn when taking regular lanes and spend when accessing the express lane. We formulated the problem as a finite-population dynamic game, used a mean-field approximation to design token prices with limited behavioral assumptions, and validated the approach on a microscopic simulation of the San Mateo US-101 corridor. The results indicate that the token economy preserves system-level efficiency: average travel time remains essentially unchanged relative to a baseline scenario in which no lane is reserved as an express lane. Furthermore, the proposed mechanism induces turn-taking among users, while accounting for users' varying urgency on their trips. Indeed, the urgency-weighted perceived travel time decreases by roughly 9\% relative to the baseline. Overall, this work suggests that token economies can provide a practical and fairer alternative to monetary tolling.
Future work should take into account the number of passengers aboard each vehicle in more detail, although specific data \rev{are} not available. Ideally, the efficiency metric would be person flow instead of vehicle flow, which would require designing token prices that depend on the number of passengers aboard, for which there are known techniques \citep{PedrosoAgazziEtAl2024EqtEql}.

\vspace{-0.4cm}

{\small \bibliography{references-c,bibliography,references-gt,references}}

\end{document}